\documentclass[
]{ceurart}

\usepackage{listings}
\usepackage{silence}
\usepackage{silence}
\usepackage{tikz}
\usetikzlibrary{arrows.meta,positioning,fit,calc,shapes.geometric,shapes.misc}
\usepackage{booktabs}

\begin{document}

\copyrightyear{2026}
\copyrightclause{Copyright for this paper by its authors.
  Use permitted under Creative Commons License Attribution 4.0
International (CC BY 4.0).}


\title{Traceable by Design: An LLM Pipeline and Dashboard for EU Regulatory Consultation Analysis}

\author[1]{Thales Bertaglia}
\author[1]{Haoyang Gui}
\author[1]{Catalina Goanta}
\author[2]{Gerasimos Spanakis}
\address[1]{Utrecht University, The Netherlands}
\address[2]{Maastricht University, The Netherlands}

\begin{abstract}
Public consultations generate large volumes of data in the form of stakeholder submissions that are practically unfeasible to analyse manually. We present an end-to-end LLM-based pipeline and interactive dashboard for structured topic extraction from regulatory consultation submissions, demonstrated on the European Commission's Digital Fairness Act (DFA) public call for evidence as a case study. The system processes raw PDF attachments and web-form responses, extracts topic annotations, and grounds every extraction in a verbatim quote from the source text. Applied to 4,322 DFA submissions, the pipeline produced 15,368 topic annotations supported by 20,951 verbatim evidence quotes. Three principles govern the proposed design: verbatim grounding, full traceability, and transparency by design. The dashboard exposes the full extraction dataset through five analytical views, from dataset-level topic overviews to individual paragraph drill-downs, with every result traceable to its source. Beyond the predefined DFA topic categories, the pipeline generated certain stakeholder concerns, such as Age Verification, Payment Processor Censorship, and Digital Ownership, that a fixed-taxonomy approach would have missed. The pipeline is domain-generic; adapting it to a new consultation requires only a prompt update and a new dataset. A live demo is available at \url{https://dfa-dashboard.thalesbertaglia.com/}. The code and processed data are publicly available at \url{https://github.com/thalesbertaglia/dfa-dashboard}.

\end{abstract}


\maketitle

\section{Introduction}
EU public consultations and calls for evidence represent a formal mechanism of evidence-based policymaking. Before major regulatory reforms are adopted, the European Commission invites citizens, businesses, civil society organisations, and academic institutions to submit written responses via the ``Have Your Say'' portal \cite{EuropeanCommissionHave2025}. These submissions constitute a form of structured public testimony: stakeholders articulating positions, raising concerns, and providing supporting arguments that regulators must take into account in the democratic decision-making process. As consultations generate large forms of data in volume and visibility, it becomes practically infeasible to analyse all submissions manually.

The public consultation on the EU Digital Fairness Act (DFA) illustrates the challenge. The DFA is expected to update the EU consumer acquis, and as such has attracted a lot of attention from the public at large. Between July and October 2025, the consultation attracted 4,325 responses covering topics such as dark patterns, addictive design, unfair personalisation, misleading influencer marketing, and exploitative subscription practices. The submissions are heterogeneous in form: 94.1\% are short web-form responses submitted directly via the portal, while the remaining 5.9\% include multi-page PDF attachments that elaborate on the feedback. Taken together, they constitute a large dataset that is challenging to systematically review at scale without computational support.

To enable systematic analysis at this scale, we present an end-to-end pipeline and interactive dashboard for structured evidence extraction from regulatory consultation submissions. The system transforms raw PDF and web-form documents into auditable evidence records: for each submission, it uses LLMs to extract the topics raised and the verbatim passages that support each topic claim. Applied to the preprocessed and filtered DFA dataset, the pipeline processed 4,322 documents, producing 15,368 topic annotations grounded in 20,951 verbatim evidence quotes. The DFA consultation serves as a case study; the pipeline is domain-generic and can be extended to other consultations and policy domains. Prior computational approaches to consultation analysis have focused primarily on document-level clustering and topic categorisation \cite{Livermore2017, di2024mining}. These methods answer aggregate questions about what topics appear across the corpus, but do not produce structured, source-grounded evidence records: they cannot identify which specific passage in which specific submission supports a given position, nor provide a traceable link from an extracted claim back to its verbatim source. Filling this gap is the central contribution of this work.

Three design principles govern every stage of the system. \textit{Provenance}: every extracted item is anchored to a verbatim quote from the source text, making the extraction verifiable against the original document. \textit{Transparency}: each processing step produces independently inspectable intermediate outputs, and pipeline metadata is surfaced directly in the dashboard interface. \textit{Traceability}: the system maintains a complete audit trail from raw input to structured record, so that any extracted claim can be traced back through the processing chain to its source paragraph. These properties are fundamental because the outputs are intended to inform policy decisions. An extracted claim that cannot be traced back to its source text provides no reliable basis for regulatory or legal reasoning \cite{bex2010hybrid, rashkin2023measuring}.

This paper describes the system and its application to the DFA consultation dataset, focusing on the pipeline architecture, interface design and practical lessons from processing a large and heterogeneous consultation dataset. Section~\ref{sec:dashboard} describes the dashboard and its main features. The paper is organised as follows. Section~\ref{sec:related} discusses related work. Section~\ref{sec:system} describes the pipeline in detail. Section~\ref{sec:limitations} discusses the limitations of the system and key takeaways from the design and development process. Section~\ref{sec:conclusion} presents our conclusions.

\section{Related Work}
\label{sec:related}
The legal NLP field has produced a substantial set of resources and benchmarks for EU regulatory text, including EURLEX57K~\cite{chalkidis2019eurlex}, LEGAL-BERT~\cite{chalkidis2020legal}, and the LexGLUE benchmark~\cite{chalkidis2022lexglue}, establishing that EU legislative documents have distinct linguistic properties that reward domain-specific modelling. Apart from laws themselves as a source of data, more recently, \citet{goanta2023} argued that legal NLP should also systematically analyse regulatory processes. This framing is adopted in our work directly: we extend the RegNLP scope from analysis of enacted regulation to structured extraction from the upstream consultation submissions that inform it. Existing legal NLP overwhelmingly targets legislation, case law or contracts \cite{Surden2014, Zhong2020, siino2025exploring}, generally leaving consultation submissions largely unexplored.

One of the earliest large-scale attempts to analyse regulatory submissions computationally is \citet{Livermore2017}, who applied topic modelling and duplicate detection to nearly three million US federal regulatory comments. \citet{romberg2024making} surveys subsequent work across multiple jurisdictions, finding that most systems rely on traditional NLP methods and explicitly calling for tools that keep human evaluators in the loop, a gap our auditability-centred design directly addresses. The closest prior work to ours is \citet{di2024mining}, who applied TF-IDF and BERT-based clustering to 830 consultation responses on the AI Act, DMA, and DSA from the same ``Have Your Say'' platform. Their approach generates document-level topic clusters, but no structured evidence records and no verbatim grounding. An earlier study applied simpler text analysis to DSA/DMA submissions~\cite{di2021see}. Among more recent systems, \citet{chenene2026stakeholder} combine actor detection, topic linking and RAG-based argument extraction for stakeholder debates, though they target media sources rather than formal regulatory submissions and lack verbatim grounding.

Argument mining and evidential reasoning provide the broader context for what structured evidence extraction should ultimately support. \citet{lawrence2019argument} survey argument component identification across domains; \citet{habernal2024mining} extend this to legal settings with the largest annotated dataset for ECtHR proceedings. For evidential reasoning specifically, \citet{bex2010hybrid} establish that structured, traceable evidence is foundational to formal reasoning over legal claims. Our pipeline outputs -- topic, stakeholder, verbatim quote and source paragraph -- constitute the evidence layer on which this kind of downstream reasoning could operate.

\section{Dashboard}
\label{sec:dashboard}
The dashboard is a web application publicly accessible at \url{https://dfa-dashboard.thalesbertaglia.com/}. It is organised around a persistent sidebar and five analytical tabs. All views respond to sidebar filter changes in real time, so every metric and evidence feed reflects the current filtered selection.

\paragraph{Sidebar: Filters and Provenance}
The sidebar serves two functions. The first is filtering: seven dimensions are available, stakeholder type, organisation name (free-text substring search), scope, governance level, company size, country, and language. All filters operate conjunctively and propagate to every tab simultaneously. The second function is provenance disclosure. A collapsible expander surfaces extraction run metadata read directly from the pipeline JSON outputs: model name, schema version, temperature, maximum tokens, and run start and finish timestamps. The sidebar also exposes a dataset version selector. Multiple labelled extraction runs can coexist in the outputs directory, each identified by a \texttt{label} file. Switching between runs updates all tabs instantly, enabling direct comparison across model versions or parameter settings.

\begin{figure*}[t]
\centering
\includegraphics[width=.9\textwidth]{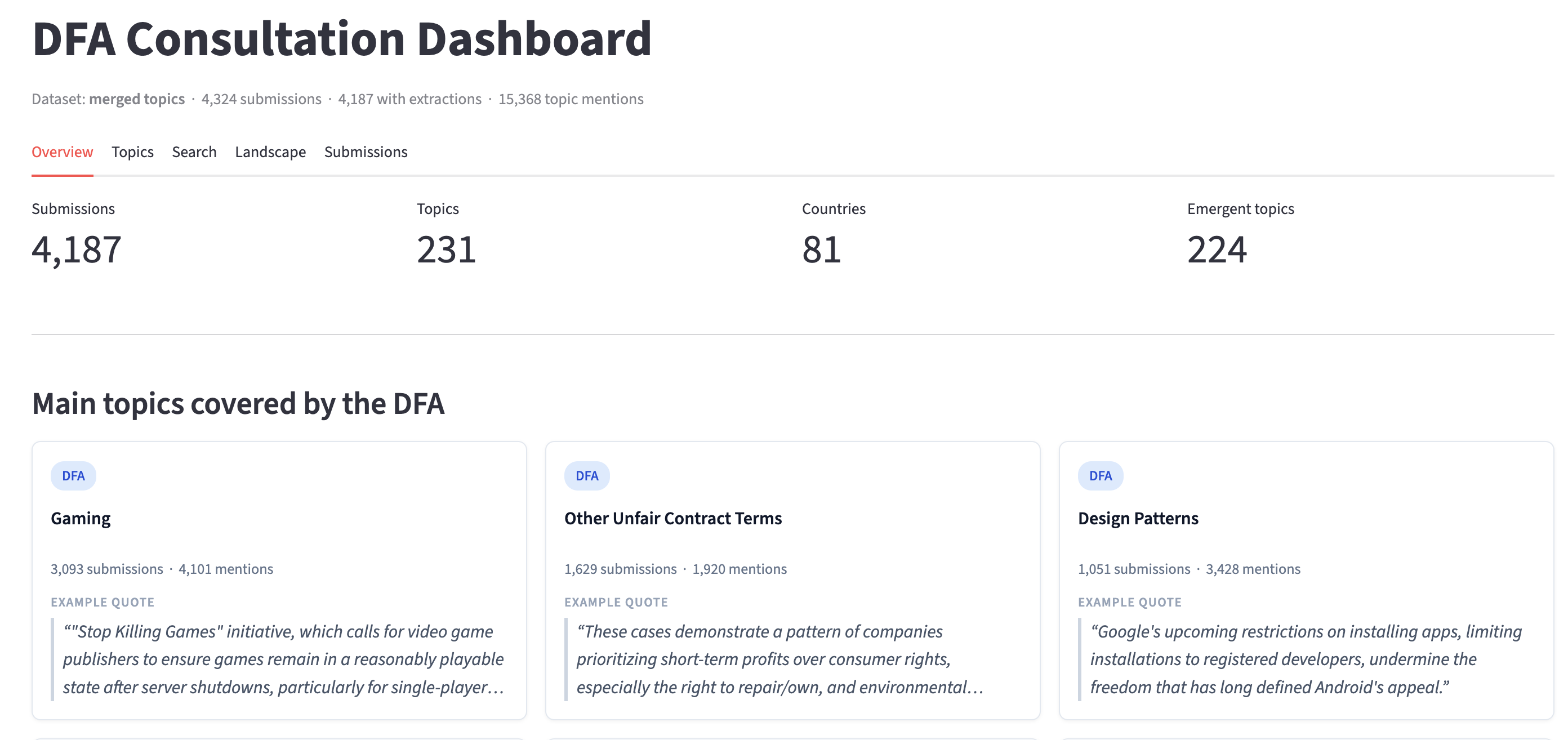}
\caption{Overview tab of the DFA Consultation Dashboard. Headline metrics (submissions, topics, countries, emergent topic count) are shown at the top. Topic cards are split into predefined DFA categories (blue) and emergent topics identified by the model (orange). Each card shows submission count, mention count, and a representative anchor quote.}
\label{fig:dashboard_overview}
\end{figure*}

\paragraph{Tabs}
The \textit{Overview} tab provides a dataset-level summary of the extraction results. Figure~\ref{fig:dashboard_overview} shows a partial view of the tab for the full DFA dataset. Four headline metrics are displayed at the top: total submissions, distinct topics, countries represented, and emergent topic count.

Below the metrics, topic cards are split into two groups. The first group covers the seven predefined DFA topic categories; the second covers emergent topics identified by the model. Each card shows the topic name, submission count, mention count, and an anchor evidence quote. DFA-registry cards are rendered in blue; emergent topic cards in orange. This colour coding is consistent across all tabs and provides the reader with an immediate visual signal about whether a topic was anticipated by the DFA's original framing or surfaced by the dataset.

The \textit{Topics} tab shifts from a dataset-level overview to topic-level evidence. Selecting any topic from a dropdown reveals all source paragraphs that mention it, grouped by document. Each paragraph is rendered with its original text alongside the evidence quotes extracted from it, displayed in cards with submitter metadata badges. A right-hand column shows the stakeholder type breakdown as a bar chart and the top ten contributing countries, enabling immediate cross-stakeholder comparison for any single topic without leaving the view. The Search tab provides free-text search across all evidence quotes in the dataset, with case-insensitive substring matching. Each result is rendered as a quote card with its topic badge and stakeholder metadata. The primary use case is targeted retrieval: finding every submission that mentions a specific product name, company, regulatory instrument, or term. 

\begin{figure*}[t]
\centering
\includegraphics[width=.8\textwidth]{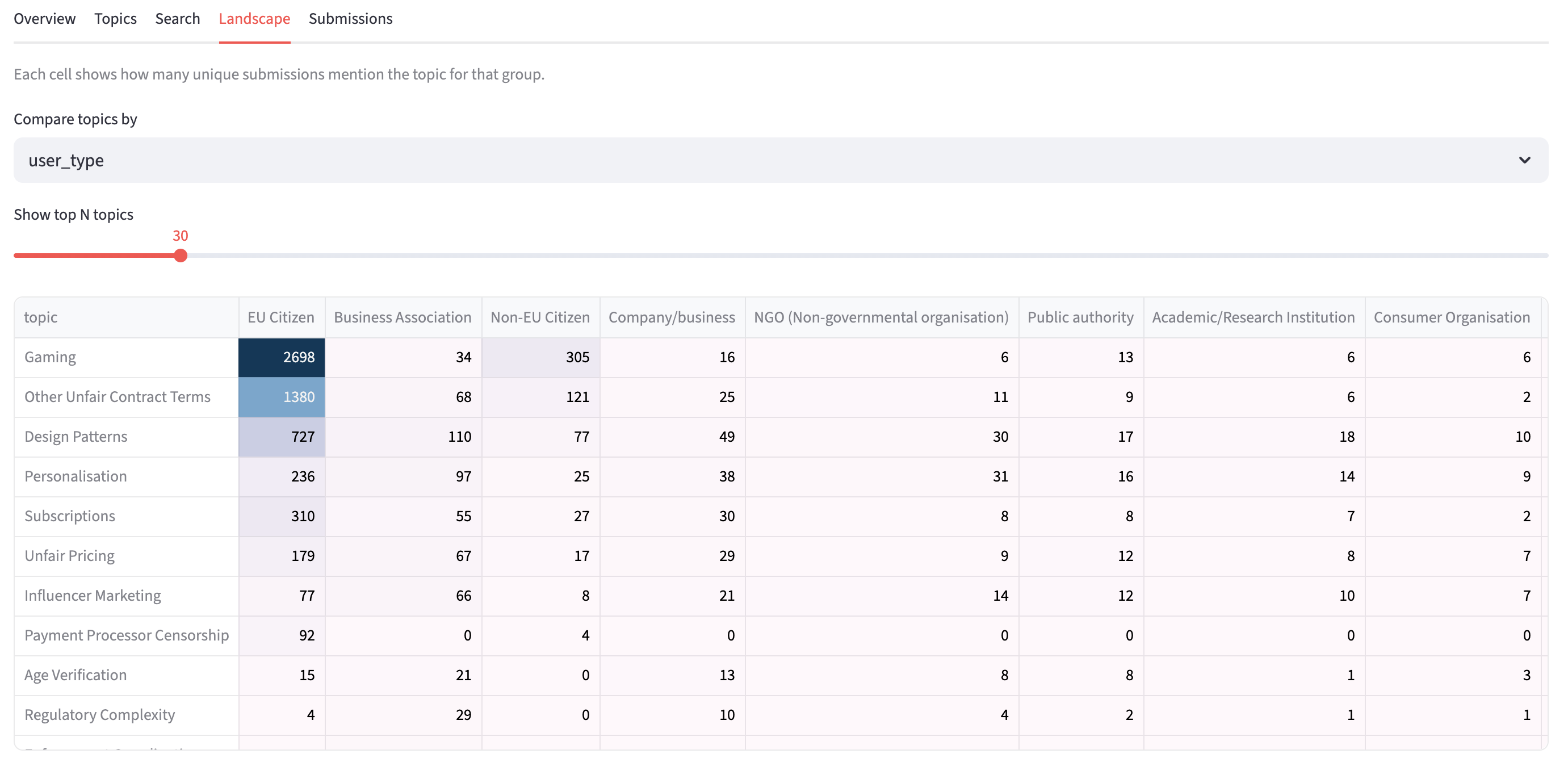}
\caption{Landscape tab configured with stakeholder type as the column dimension.}
\label{fig:dashboard_landscape}
\end{figure*}

The \textit{Landscape} tab is the dashboard's main feature for policy analysis. Figure~\ref{fig:dashboard_landscape} shows the tab configured with stakeholder type as the column dimension. The view presents a topic-by-dimension pivot table: rows are topics (the top N by submission count, configurable via a slider with a default of 30), columns are any of six dimensions (stakeholder type, country, governance level, company size, language, or scope), and cell values are unique submission counts rendered with a heatmap colour scale. The dimension selector makes the same extraction data re-addressable from multiple analytical angles without any reprocessing. From any cell, the user can focus on the evidence quotes from exactly those submissions for that topic-dimension intersection. Selecting, for example, the cell at \textit{Design Patterns} $\times$ \textit{Business Association} surfaces the passages from business association submissions that discuss design patterns.

Finally, the \textit{Submissions} tab allows users to view any individual submission in full, along with its extracted evidence. A searchable dropdown lists all submissions that match the current filters, each labelled by document ID, organisation, country, language, and date. 

\section{System Overview}
\label{sec:system}
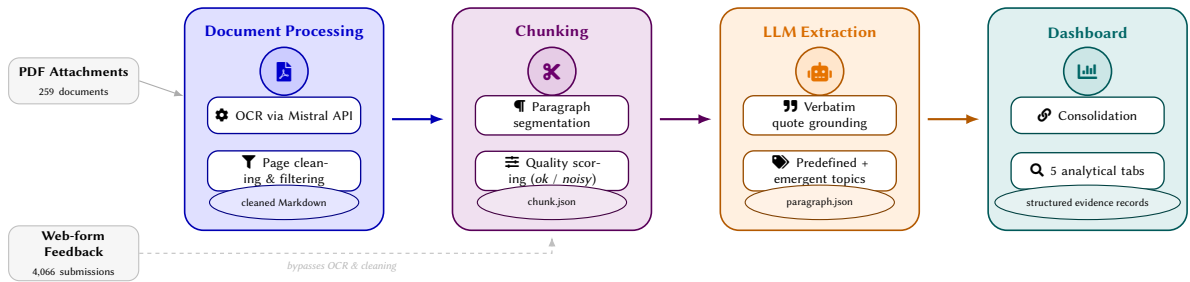
\begin{figure*}[t]
    \centering
    \resizebox{0.98\textwidth}{!}{%
    \begin{tikzpicture}[
        font=\small,
        >=Latex,
        x=1cm,
        y=1cm,
        stepbox/.style={
            rounded corners=14pt,
            very thick,
            draw=#1!70!black,
            fill=#1!12,
            minimum width=5.2cm,
            minimum height=5.8cm,
            align=center
        },
        inputbox/.style={
            rounded corners=8pt,
            thick,
            draw=gray!50,
            fill=gray!7,
            minimum width=3.4cm,
            minimum height=1.2cm,
            align=center,
            text width=3.0cm
        },
        iconbubble/.style={
            circle,
            draw=#1!70!black,
            fill=#1!20,
            very thick,
            minimum size=1.25cm,
            inner sep=0pt
        },
        mini/.style={
            rounded corners=8pt,
            draw=#1!60!black,
            fill=white,
            line width=1pt,
            minimum width=4.0cm,
            minimum height=0.95cm,
            align=center,
            text width=3.6cm
        },
        flow/.style={-{Latex[length=3.5mm]}, ultra thick, draw=#1!70!black},
        inputflow/.style={-{Latex[length=2.5mm]}, thick, draw=gray!65},
        bypass/.style={-{Latex[length=2.5mm]}, thick, dashed, draw=gray!50},
        glow/.style={ellipse, draw=#1!50!black, fill=#1!10, thick}
    ]

    \node[inputbox] (pdf) at (-5.5, 1.0) {
        \textbf{PDF Attachments}\\[2pt]
        {\scriptsize 259 documents}
    };
    \node[inputbox] (fb) at (-5.5, -3.5) {
        \textbf{Web-form Feedback}\\[2pt]
        {\scriptsize 4,066 submissions}
    };

    \node[stepbox=blue]   (b1) at (0,    0) {};
    \node[stepbox=violet] (b2) at (7.0,  0) {};
    \node[stepbox=orange] (b3) at (14.0, 0) {};
    \node[stepbox=teal]   (b4) at (21.0, 0) {};

    \node[font=\bfseries\large, text=blue!80!black]
        at ($(b1.north)+(0,-0.65)$) {Document Processing};
    \node[iconbubble=blue] at ($(b1.center)+(0,1.25)$)
        {\Large\textcolor{blue!80!black}{\faFilePdf}};
    \node[mini=blue] at ($(b1.center)+(0,0.1)$)
        {\faCog\enspace OCR via Mistral API};
    \node[mini=blue] at ($(b1.center)+(0,-1.35)$)
        {\faFilter\enspace Page cleaning \& filtering};
    \node[glow=blue, minimum width=3.9cm, minimum height=0.9cm]
        at ($(b1.south)+(0,0.75)$)
        {\scriptsize cleaned Markdown};

    \node[font=\bfseries\large, text=violet!80!black]
        at ($(b2.north)+(0,-0.65)$) {Chunking};
    \node[iconbubble=violet] at ($(b2.center)+(0,1.25)$)
        {\Large\textcolor{violet!85!black}{\faCut}};
    \node[mini=violet] at ($(b2.center)+(0,0.1)$)
        {\faParagraph\enspace Paragraph segmentation};
    \node[mini=violet] at ($(b2.center)+(0,-1.35)$)
        {\faSlidersH\enspace Quality scoring (\textit{ok} / \textit{noisy})};
    \node[glow=violet, minimum width=3.9cm, minimum height=0.9cm]
        at ($(b2.south)+(0,0.75)$)
        {\scriptsize chunk.json};

    \node[font=\bfseries\large, text=orange!85!black]
        at ($(b3.north)+(0,-0.65)$) {LLM Extraction};
    \node[iconbubble=orange] at ($(b3.center)+(0,1.25)$)
        {\Large\textcolor{orange!90!black}{\faRobot}};
    \node[mini=orange] at ($(b3.center)+(0,0.1)$)
        {\faQuoteRight\enspace Verbatim quote grounding};
    \node[mini=orange] at ($(b3.center)+(0,-1.35)$)
        {\faTags\enspace Predefined + emergent topics};
    \node[glow=orange, minimum width=3.9cm, minimum height=0.9cm]
        at ($(b3.south)+(0,0.75)$)
        {\scriptsize paragraph.json};

    \node[font=\bfseries\large, text=teal!80!black]
        at ($(b4.north)+(0,-0.65)$) {Dashboard};
    \node[iconbubble=teal] at ($(b4.center)+(0,1.25)$)
        {\Large\textcolor{teal!80!black}{\faChartBar}};
    \node[mini=teal] at ($(b4.center)+(0,0.1)$)
        {\faLink\enspace Consolidation};
    \node[mini=teal] at ($(b4.center)+(0,-1.35)$)
        {\faSearch\enspace 5 analytical tabs};
    \node[glow=teal, minimum width=3.9cm, minimum height=0.9cm]
        at ($(b4.south)+(0,0.75)$)
        {\scriptsize structured evidence records};

    \draw[flow=blue]   ($(b1.east)+(0.2,0)$)  -- ($(b2.west)+(-0.2,0)$);
    \draw[flow=violet] ($(b2.east)+(0.2,0)$)  -- ($(b3.west)+(-0.2,0)$);
    \draw[flow=orange] ($(b3.east)+(0.2,0)$)  -- ($(b4.west)+(-0.2,0)$);

    \draw[inputflow] (pdf.east) -- ($(b1.west)+(0,0.6)$);

    \draw[bypass] (fb.east)
        -- (7.0, -3.5)
        -- ($(b2.south)+(0,-0.15)$);

    \node[font=\scriptsize\itshape, text=gray!55, align=center]
        at (1.5, -3.85) {bypasses OCR \& cleaning};

    \end{tikzpicture}%
    }
    \caption{Pipeline architecture. PDF attachments (259 documents) and web-form feedback texts (4,066 submissions) follow distinct ingestion paths. PDFs are converted via OCR, cleaned page-by-page, and segmented into paragraph-level units. Feedback texts bypass these stages and are ingested directly as single units. Both paths converge at the chunking stage, where quality labels (\textit{ok}\,/\,\textit{noisy}) gate LLM access. The extraction stage produces verbatim-grounded topic annotations, which are consolidated with source text and surfaced through the interactive dashboard.}
    \label{fig:pipeline_architecture}
\end{figure*}

The system transforms raw consultation submissions into structured, source-grounded evidence records and exposes the resulting data through an interactive dashboard designed for policy analysis. Figure~\ref{fig:pipeline_architecture} shows the end-to-end architecture. Two input paths feed a shared processing pipeline: PDF attachments are converted via OCR and segmented into paragraph-level units, while web-form feedback texts are ingested directly as single units. Both paths produce chunk JSON files that feed the LLM extraction stage. Extraction outputs are then consolidated with their source text and loaded into the dashboard.

Three principles govern every design decision in the system. \textit{Verbatim grounding}: no paraphrase or summary is introduced at any stage; every extracted topic is anchored to a verbatim quote copied exactly from the source text. \textit{Full traceability}: every item visible in the dashboard has an unbroken chain of evidence from the displayed evidence quote back to the source paragraph, source document and original submission. \textit{Transparency by design}: no stage overwrites its input; every intermediate output is independently inspectable, and pipeline metadata is surfaced directly in the dashboard interface so that any result can be traced to the exact configuration that produced it.

The system is designed to be domain-generic. The OCR, cleaning, chunking, and LLM extraction components operate on any PDF or text corpus. The only domain-specific element is the topic schema embedded in the extraction prompt, which comprises seven predefined DFA issue categories with mapping instructions. Replacing this schema with another requires modifying a single prompt file. The DFA Consultation is one case-study of how the architecture can be used.

\subsection{Document Processing Pipeline}

The document processing pipeline transforms raw source documents into paragraph-level units ready for LLM extraction. Four stages handle ingestion, optical character recognition (OCR, to extract text from the PDF attachments), cleaning and filtering and chunking with quality scoring. Each stage writes named intermediate outputs rather than passing data in memory, so any stage can be re-run, inspected, or replaced independently, directly implementing our transparency principle.

\paragraph{Dataset} 
The source dataset consists of submissions to the European Commission's ``Have Your Say'' platform for the DFA consultation, received between July and October 2025. The full submission set comprises 4,325 entries, of which 3 were excluded for being unavailable or empty. Of the remaining, 338 included PDF attachments  (position papers and structured responses from institutional stakeholders) and the rest were text-only web-form responses. The PDF attachments were manually filtered to excluded non-submission files such as appended academic papers or annex-only documents, resulting in a final set of 259 documents. These two input types are handled differently downstream, as described below.

\paragraph{OCR}
PDF conversion is performed using the Mistral OCR API\footnote{\url{https://mistral.ai/news/mistral-ocr}}, selected for its handling of multi-column layouts and image-heavy documents that cause reading-order failures in rule-based tools. The API produces two outputs per document: a structured JSON representation and a per-page Markdown file. Both are retained as intermediate outputs alongside a \texttt{report.json} summarising conversion quality across the batch.

\paragraph{Cleaning and Filtering}
Cleaning operates on the OCR JSON output page by page. A sequential filter pipeline removes image placeholders, Markdown footnote definitions, isolated page numbers, and Unicode footnote markers, then normalises whitespace. Beyond these standard artefact removals, the pipeline applies three additional quality checks. First, automatic header and footer detection: lines that appear in more than 60\% of pages across a document are identified as boilerplate and stripped. Second, lexical diversity checking: pages where the ratio of unique to total words falls below 0.15 are flagged as likely garbled OCR output and excluded. Third, table-of-contents detection: pages matching structural TOC patterns are removed as non-substantive. Pages falling below a 30-word threshold after cleaning are also excluded. Documents are then bucketed based on the number of pages dropped: documents losing two or fewer pages are written to \texttt{reliable/}; those losing more are written to \texttt{flag/}. Flagged documents are still processed downstream, but the flag indicates that extraction coverage should be interpreted with caution.

\paragraph{Chunking and Quality Scoring}
The goal of chunking is to segment cleaned documents into semantically coherent units sufficient for LLM extraction, which provide enough context to support verbatim quote extraction without diluting the topical signal. We parse cleaned Markdown using \texttt{markdown-it-py}\footnote{\url{https://github.com/executablebooks/markdown-it-py}}, extracting paragraphs and headings, and list items with their section context preserved. Two input types are handled differently at this stage, distinguished by the \texttt{unit\_kind} field. \texttt{paragraph} units are produced by the Markdown parser from PDF-derived documents and go through quality scoring. \texttt{feedback} units are the raw web-form text from the dataset CSV, ingested as a single self-contained unit per submission and assigned \texttt{chunk\_quality=``ok''} directly, bypassing scoring entirely. This reflects the nature of web-form responses: they are already the minimal meaningful unit and require no further segmentation.

For \texttt{paragraph} units, each chunk receives a meaning score: $\text{score} = 0.55 \times \ell + 0.45 \times \alpha - 0.35 \times \delta$ where $\ell$ is a length component, $\alpha$ an alpha-character ratio component, and $\delta$ a digit penalty, all clamped to $[0,1]$. A hard floor applies: paragraphs with an alpha-character ratio below $0.35$ receive a score of $0.0$ regardless of other components. Chunks scoring at or above the default threshold of $0.50$ are labelled \texttt{chunk\_quality=``ok''}; a top-$k$ fallback (default $k=60$) ensures minimum coverage even for short documents. Chunks labelled \texttt{``noisy''} are skipped by the LLM but recorded in the output, thus preserving a full account of what the model did not see. The output of this stage is a \texttt{<doc\_id>.chunk.json} file per document containing paragraph text, section path, meaning score, selection flag and \texttt{unit\_kind} for every chunk. This file is the primary auditing artefact for the extraction stage, as it allows a user to verify exactly which paragraphs of a given document were and were not presented to the model.

\subsection{LLM-Based Extraction}
LLM calls are managed via LiteLLM\footnote{\url{https://litellm.ai}}, which abstracts over model providers and makes the pipeline model-agnostic. We use \textit{gpt-5-nano} with temperature 1.0. Only chunks with \texttt{chunk\_quality=``ok''} are sent to the model; skipped chunks are recorded with a \texttt{skip\_reason} field. 

\paragraph{Prompt Design}

The extraction prompt includes the paragraph text as its only variable input. The DFA topic schema (seven predefined categories with mapping instructions) is embedded directly in the prompt, along with explicit rules about extraction behaviour. The prompt instructs the model to extract only topics explicitly supported by the text, not to infer or generalise beyond what is written, and to return empty output for passages consisting of names, job titles, affiliations, signatures, headers, or procedural text. We hypothesise that these negative rules contribute to the 33.6\% empty-output rate across chunks sent to the model.

The verbatim grounding rules are also included in the prompt: every extracted topic must include at least one evidence quote copied exactly from the passage, with no paraphrasing, no shortening, no normalisation, and no correction. The prompt specifies the smallest exact quote that clearly supports the topic, which may be a full sentence or a supporting fragment, but must appear verbatim in the source. We enforce the implementation of the provenance principle using a Pydantic schema requiring at least one non-empty string in the \texttt{evidence\_quotes} field, thus ensuring that the models always outputs a quote. We verified verbatim grounding by matching all extracted quotes against their source paragraphs using exact and fuzzy string matching, achieving an overall match rate of 99.1\%.

Each extracted item carries a \texttt{topic\_source} flag: \texttt{``dfa''} for topics mapped to one of the seven predefined DFA categories, and \texttt{``new''} for topics the model identifies outside that schema. The prompt instructs the model to prefer predefined labels when they fit and to use new labels only when no predefined category reasonably applies, and to make new labels short, specific and concrete rather than generic abstractions. This hybrid design ensures coverage of the issues the DFA was designed to address while remaining open to concerns stakeholders raise beyond that framing. In the DFA run, 88.6\% of extracted items mapped to predefined labels; 11.4\% were emergent, producing 1,458 unique raw emergent labels. Many of the generated emergent topics are near-duplicate or redundant labels, so we perform a post-processing step to increase their quality.

\paragraph{Topic Post-Processing}
The 1,458 unique emergent labels produced by the raw extraction run require consolidation before they are analytically usable. A three-step post-processing pipeline addresses this. First, emergent labels are mapped to the seven predefined DFA topics using semantic similarity, and unmatched labels are passed to the next step. Second, remaining labels are grouped via agglomerative clustering; each cluster receives a canonical label selected by frequency and brevity. Third, residual low-frequency labels are matched to the nearest existing group via similarity and reviewed manually before final assignment. This procedure reduces the emergent label space from 1,458 to 224 distinct topics while preserving substantively meaningful niche concerns that would otherwise be lost. The dashboard exposes both the raw and post-processed topic sets, letting analysts inspect the effect of consolidation directly.

Finally, the consolidation step re-joins LLM extraction records with their source paragraph text from the chunk files. For each extracted paragraph record, the script locates the matching chunk by \texttt{para\_id} and attaches the original text alongside optional chunk metadata. The final consolidated record carries: \texttt{doc\_id}, \texttt{para\_id}, \texttt{topic}, \texttt{topic\_source}, \texttt{evidence\_quotes}, and the source paragraph text. This results in a complete traceability chain -- from any item in the dashboard, a user can follow this chain back to the exact text the model was shown.

\section{Limitations and Practical Considerations}
\label{sec:limitations}
Building and deploying the pipeline on a real consultation dataset surfaces observations not visible from the architecture alone. This section documents what we found in practice: where the design held up, where it created friction, and what we would recommend to anyone applying a similar approach.
\paragraph{Transparency as a design requirement} Every pipeline parameter shapes what gets extracted: the meaning score threshold determines which paragraphs the model sees; the topic schema determines which issues are predefined versus emergent; the prompt's negative rules determine what counts as substantive content. These are interpretive choices with real consequences in a policy context. Our recommendation is to treat transparency as a first-class requirement from the start. Concretely: retain intermediate outputs at every stage, version the prompt alongside the code, and surface pipeline configuration directly in any interface that presents results. The dashboard's provenance expander operationalises this principle.

\paragraph{OCR and chunking are the real bottlenecks} LLM extraction quality is bounded by text quality, which is set upstream by OCR and chunking. These steps receive far less attention than prompt engineering in most LLM pipeline discussions, but in practice, they are where the most significant and least recoverable errors occur. OCR failures, such as reading-order errors in multi-column layouts, garbled tables, misidentified page boundaries, corrupt the text units that all subsequent steps operate on. No prompt refinement recovers a position paper whose columns were read across rather than down. Manual sampling of a representative subset of documents before running the full pipeline is worth the time. The \texttt{flag/} bucketing mechanism surfaces problematic documents but does not fix them. Chunking introduces a second layer of loss: the meaning score heuristic handles standard narrative prose well but degrades on tables, itemised policy recommendations, and short formulaic statements that carry substantive content. We recommend treating the threshold as a tunable parameter and spot-checking excluded chunks on a sample of documents. LLM-assisted segmentation by content structure rather than typographic convention is the most promising longer-term improvement.

\paragraph{Emergent labels: a useful tradeoff with a real cost} The raw extraction produced 1,458 unique emergent topic labels, consolidated to 224 after post-processing. Near-duplicate labels(such as \textit{Payment Processor Censorship}, \textit{Payment Processor Gatekeeping}, \textit{Payment Processor Pressure}) are a direct consequence of processing each paragraph independently without a shared vocabulary across documents. This is a known tradeoff: the same independence that generates near-duplicates also surfaces genuine niche concerns that a more constrained approach would suppress, including policy signals such as Age Verification, Digital Ownership, and Enforcement Coordination that a fixed taxonomy would miss entirely. The practical implication is to plan for post-processing from the start, treat the cluster distance threshold as a parameter requiring calibration per dataset, and build a manual review step into the workflow for low-frequency labels. The dashboard exposes both raw and merged topic sets so users can inspect the consolidation rather than accept it uncritically.

\paragraph{Verbatim grounding}
The verbatim quote constraint is enforced through prompting and schema validation. Post-hoc verification via exact and fuzzy string matching against source paragraphs confirmed a 99.1\% match rate across all extracted quotes. There are still some quotes that do not match the source text, indicating potential model hallucinations that should be further analysed.

\section{Conclusion}
\label{sec:conclusion}
We presented an end-to-end NLP pipeline and interactive dashboard for structured evidence extraction from regulatory public consultation submissions, applied to the EU Digital Fairness Act dataset as a case study. The system processed 4,322 submissions and produced 15,368 topic annotations, grounded in 20,951 verbatim evidence quotes. Three principles govern the design throughout: verbatim grounding, full traceability, and transparency by design.

The DFA application demonstrates the system's practical value. The hybrid predefined-plus-emergent topic schema surfaced policy signals outside the DFA's original framing (including Age Verification, Payment Processor Censorship, and Digital Ownership) that a fixed-taxonomy approach would have missed entirely. The Landscape tab makes cross-stakeholder analysis tractable at scale while preserving a direct path from aggregate pattern to verbatim source text. The provenance expander ensures every result is traceable to a specific, reproducible pipeline configuration.

The pipeline is domain-generic and designed for reuse. Adapting it to a new consultation requires updating the extraction prompt and supplying new data; the pipeline architecture, dashboard, and output format require no modification. As regulatory consultations grow in volume and policy significance across EU digital governance, tools that make large-scale stakeholder testimony systematically accessible while remaining faithful to source material are increasingly necessary. This system is a concrete step in that direction.

Future work include the evaluation of other models (such as open-source LLMs), LLM-assisted chunking by argumentative unit, analysis of the emergent topic consolidation procedure across multiple consultation datasets, and systematic evaluation of the quality of the extracted topics. Pipeline code and DFA extraction outputs are publicly available at \url{https://github.com/thalesbertaglia/dfa-dashboard} to enable replication and extension.

\section{Acknowledgments}
This research has been supported by funding from the ERC Starting Grant HUMANads (ERC-2021-StG No 101041824).

\bibliography{reference}

\end{document}